\documentclass[aps,preprint]{revtex4}%
\usepackage{amsfonts}
\usepackage{amsmath}
\usepackage{amssymb}
\usepackage{psfrag}
\usepackage[dvips]{graphicx}

\providecommand{\U}[1]{\protect\rule{.1in}{.1in}}
\begin{document}
\preprint{ }

\begin{center}
\qquad
%\hfill CERN-PH-TH/2010-082\\[0pt]
%\vskip -.1 cm \hfill LMU-ASC 26/10 \vskip -.1 cm \hfill MPP-2010-49\\[0pt]

{\Large \textbf{Classical Dimensional Transmutation and Confinement}} \
%\vspace{2cm}

\vspace{1cm}

\textbf{Gia Dvali}$^{a,b,d,c}$, \textbf{Cesar Gomez}$^{e}$ and \textbf{Slava
Mukhanov}$^{a,b}$

\vspace{.6truecm}

\vspace{.2truecm}

\emph{$^{a}$ASC, Department f\"{u}r Physik, LMU, M\"{u}nchen\\[0pt]%
Theresienstr.~37, 80333 M\"{u}nchen, Germany}

%\vspace{.2truecm}

\emph{$^{b}$MPI f\"{u}r Physik\\[0pt]F\"{o}hringer Ring 6, 80805 M\"{u}nchen,
Germany}

%\vspace{.6truecm}

\emph{$^{c}$CERN, Theory Division\\[0pt]1211 Geneva 23, Switzerland}

%\vspace{.2truecm}

\emph{$^{d}$CCPP, Department of Physics, NYU\\[0pt]4 Washington Place, New
York, NY 10003, USA}

\vspace{.2truecm}

\emph{$^{e}$Instituto de F\'{\i}sica Te\'orica UAM-CSIC, C-XVI \\[0pt]%
Universidad Aut\'onoma de Madrid, Cantoblanco, 28049 Madrid, Spain}\\[0pt]
\end{center}

%\vskip .4in

\centerline{\bf Abstract}
%\vskip .1in
%\no
\noindent
We observe that probing certain classical field theories by external sources
uncovers the underlying renormalization group structure, including the
phenomenon of dimensional transmutation, at purely-classical level. We perform
this study on an example of $\lambda\phi^{4}$ theory and unravel asymptotic
freedom and triviality for negative and positives signs of $\lambda$
respectively. We derive exact classical $\beta$ function equation. Solving
this equation we find that an isolated source has an infinite energy and
therefore cannot exist as an asymptotic state. On the other hand a dipole,
built out of two opposite charges, has finite positive energy. At large
separation the interaction potential between these two charges grows
indefinitely as a distance in power one third.

\vskip .4in \noindent

%\end{center}

\newpage{}
\section{Introduction}
The discovery of asymptotic freedom in QCD \cite{Gross} opened a new era in
particle physics. Besides its direct relevance for understanding the nature of
strong interactions, it introduced a concept of dimensional transmutation or
equivalently a dynamical generation of scale in a seemingly scale-free theory.
Both phenomena, asymptotic freedom and dimensional transmutation, are usually
perceived as intrinsically quantum phenomena, as they are both deeply rooted
in the renormalization group properties of the quantum theory. The question we
would like to address in this paper is, how much of these phenomena is
captured by classical physics? This is a fully legitimate question, since
usually quantum effects have classical precursors, which sometimes appear in
the form of uncontrollable growth of the classical fields.

In order to illustrate our ideas, we will consider a simple example, namely,
scalar theory with negative $\lambda\phi^{4}$, which is known to be
renormalizable and has a negative $\beta$ function \cite{simanzek}, but has an
unbounded from below potential. However, our classical renormalization group
treatment delivers a natural prescription which allows to self-consistently
work with this theory in the presence of external sources, and isolate our
findings from the issue of potential instability in a pure $\lambda\phi^{4}$ theory.

In the presence of the external sources, the requirement of independence of
physical observable from the regulator scale of sources implies the running of
the effective coupling. This running is the main reason behind the whole
renormalization group structure and the subsequent classical dimensional
transmutation. This scale dependence is the key to why the renormalization
group results can be safely disentangled from the instability issues. Simply
speaking, because of the emerging scale-dependence we always perform
calculations on time-scales shorter than would-be instability time in a
sourceless theory.

Surprisingly, by probing $\lambda\phi^{4}$ theory by the large external
source, we uncover the whole built-in RG structure already at the classical
level, with fully-fledged counterparts of asymptotic freedom as well as
dimensional transmutation phenomena, in which an analog of QCD-scale appears
as a result of \textit{classical RG invariance}. We derive the exact classical
$\beta$ function equation from which we extract non perturbative information
about the infrared region. Solving this equation in strong coupling regime we
find that the energy of the isolated external charge is infinite and
\textit{positive} and hence it cannot exist as a free asymptotic state.
Moreover, considering a dipole, build out of two charges with opposite signs,
we find that its energy is positive and finite. When charges in this dipole
are separated their interaction potential grows indefinitely as distance in
power one third, thus confining the charges. These findings indicate that
there may exist a classical counterpart of confinement.

\section{Classical solution}
After the discovery of asymptotic freedom in the non-Abelian Yang Mills
theories, the physics underlying the negative sign of the $\beta$ function was
understood as an anti-screening effect due to self-interactions of the gauge
fields. From the classical point of view we can try to understand this
anti-screening considering how self-interaction modifies the field created by
an external point-like source $Q$ at large distances. This modification
compared to the case of free fields can be used to define the effective charge
$Q_{eff}(r)$ at distance $r$ or, equivalently, the running coupling constant
$\alpha\left(  r\right)  $. This can be done without invoking quantum theory
and the result will only depend on the particular classical features of
self-interactions. For the large external charge one can expect that the
classical contribution to anti-screening will dominate over the one due to the
vacuum polarization effects.

In this section we will solve perturbatively the classical equations of motion
for a given external charge and show how the anti-screening effect (the growth
of $Q_{eff}(r)$ with $r)$ can naturally be achieved.

\subsection{Anti-screening}
Let us consider $\lambda_{0}\phi^{4}$ theory with negative $\lambda_{0}$
\begin{equation}
S=\int\left(  \frac{1}{2}\partial_{\mu}\phi\partial^{\mu}\phi-\frac{1}%
{4}\lambda_{0}\phi^{4}+4\pi Q\phi\right)  d^{4}x, \label{1}%
\end{equation}
where the signature is taken to be $+,---,$ and $Q$ is the external charge. In
the case of a point-like charge the field equation for the static spherically
symmetric field $\phi\left(  r\right)  $ reduces to%
\begin{equation}
\frac{1}{r^{2}}\frac{d}{dr}\left(  r^{2}\frac{d\phi}{dr}\right)  -\lambda
_{0}\phi^{3}=-4\pi Q\delta\left(  \mathbf{x}\right)  . \label{2}%
\end{equation}
If one neglects the nonlinear term in this equation then its solution is%
\begin{equation}
\phi_{0}=\frac{Q}{r}. \label{3}%
\end{equation}
Notice that for $Q\gg1$ the amplitude of the classical scalar field on scales
$r$ is much larger than the typical amplitude of the quantum fluctuations,
which is of order $1/r.$ If the coupling constant $\lambda_{0}$ is small
enough, i.e. $\lambda_{0}Q^{2}\ll1,$ then the corrections to solution
(\ref{2}) due to the self-interaction $\lambda_{0}\phi^{3}$ can be treated
perturbatively. The leading order correction to the solution $\phi_{0}$ can be
obtained by solving equation%
\begin{equation}
\frac{1}{r^{2}}\frac{d}{dr}\left(  r^{2}\frac{d\phi}{dr}\right)  =-4\pi
Q\delta\left(  \mathbf{x}\right)  +\lambda_{0}\frac{Q^{3}}{r^{3}}, \label{4}%
\end{equation}
where we have substituted $\phi_{0}$ in the nonlinear term. The last term in
this equation can be treated as the contribution to the effective charge
induced by the nonlinear self-interaction. As it was noticed above, for
$Q\gg1,$ the vacuum polarization contribution to the induced charge is much
smaller than the classical contribution and therefore can be neglected. As one
can easily see from (\ref{4}) the induced charge for positive $\lambda_{0}$
has a sign which is opposite to the sign of the source and the nonlinear
interaction leads to screening$.$ For negative $\lambda_{0}$ the charges have
the same sign and we have an anti-screening effect similar to the one of the
non-Abelian gauge theories. Since we are mainly interested in asymptotic
freedom and confinement we will consider only the case of negative
$\lambda_{0}.$

\subsection{The perturbative expansion}
Let us look for the solution to equation (\ref{2}) in the following form%
\begin{equation}
\phi\left(  r\right)  =\frac{Qf\left(  r\right)  }{r}. \label{5a}%
\end{equation}
Integrating equation (\ref{2}) and substituting this ansatz we can rewrite the
equation for the scalar field as%
\begin{equation}
f\left(  r\right)  =1+\alpha_{0}r\int_{r}^{\infty}\left(  \int_{r_{0}%
}^{r^{\prime}}\frac{f^{3}\left(  r^{\prime\prime}\right)  }{r^{\prime\prime}%
}dr^{\prime\prime}\right)  \frac{dr^{\prime}}{r^{\prime2}}-N\left(  \alpha
_{0}\right)  , \label{6a}%
\end{equation}
where
\begin{equation}
\alpha_{0}\equiv-\lambda_{0}Q^{2}>0, \label{7a}%
\end{equation}
is the effective coupling constant and we have introduced the ultraviolet
cutoff scale $r_{0}$ to regularize the integral, which otherwise would
diverge. The function $N\left(  \alpha_{0}\right)  ,$ which depends only on
$\alpha_{0}$, is fixed by the normalization condition: $f\left(  r_{0}\right)
=1.$ It is clear that in the limit $\alpha_{0}\rightarrow0$ it must vanish and
therefore in the absence of self-interaction, the solution $\left(
\ref{5a}\right)  $ with $f\left(  r\right)  =1$ exactly satisfies equation
$\left(  \ref{2}\right)  $. As it follows from (\ref{5a}), the function $f$
defines the anti-screened effective charge $Q_{eff}\left(  r\right)
=Qf\left(  r\right)  $ or, equivalently, the running coupling constant%
\begin{equation}
\alpha_{eff}\left(  r\right)  =\alpha_{0}f^{2}\left(  r\right)  \label{8a}%
\end{equation}
Assuming that $\alpha_{0}\ll1$ we can solve the integral equation $\left(
\ref{7a}\right)  $ by iterations in powers of $\alpha_{0}.$ With this purpose
it is convenient to rewrite it as%
\begin{equation}
f\left(  x\right)  =1+\alpha_{0}e^{x}\int_{x}^{\infty}\left(  \int
_{0}^{x^{\prime}}f^{3}\left(  x^{\prime\prime}\right)  dx^{\prime\prime
}\right)  e^{-x^{\prime}}dx^{\prime}-N\left(  \alpha_{0}\right)  , \label{9a}%
\end{equation}
where we have introduced $x=\ln\left(  r/r_{0}\right)  $ instead of $r.$
Substituting $f\left(  x\right)  =1$ into the right hand side of equation
$\left(  \ref{9a}\right)  $ and taking into account that $N\left(  \alpha
_{0}\right)  =\alpha_{0}+O\left(  \alpha_{0}^{2}\right)  $ we find%
\begin{equation}
f\left(  x\right)  =1+\alpha_{0}x+O\left(  \alpha_{0}^{2}\right)  .
\label{10a}%
\end{equation}
Next we take this solution, substitute it again in $\left(  \ref{9a}\right)  $
and take $N\left(  \alpha_{0}\right)  =\alpha_{0}+3\alpha_{0}^{2}+O\left(
\alpha_{0}^{3}\right)  .$ Keeping only the terms up to second order in
$\alpha_{0}^{2}$ leads to%
\begin{equation}
f\left(  x\right)  =1+\alpha_{0}x+\alpha_{0}^{2}\left(  \frac{3}{2}%
x^{2}+3x\right)  +O\left(  \alpha_{0}^{3}\right)  . \label{11a}%
\end{equation}
This procedure can be iterated giving us at each step the next order term in
$\alpha_{0}$. The result up to order $\alpha_{0}^{6}$ is%
\begin{align}
f\left(  x\right)     =&1+\alpha_{0}x+\alpha_{0}^{2}\left(  \frac{3}{2}%
x^{2}+3x\right)  +\alpha_{0}^{3}\left(  \frac{5}{2}x^{3}+12x^{2}+24x\right)\nonumber\\
&+\alpha_{0}^{4}\left(  \frac{35}{8}x^{4}+\frac{71}{2}x^{3}+\frac{285}{2}%
x^{2}+285x\right) \nonumber\\
&  +\alpha_{0}^{5}\left(  \frac{63}{8}x^{5}+93x^{4}+\frac{1143}{2}%
x^{3}+2142x^{2}+4284x\right) \nonumber\\
&  +\alpha_{0}^{6}\left(  \frac{231}{16}x^{6}+\frac{9129}{40}x^{5}+\frac
{7665}{4}x^{4}+10\,521x^{3}+37\,989x^{2}+75\,978x\right)  +O\left(  \alpha
_{0}^{7}\right)  . \label{12a}%
\end{align}
The function $N\left(  \alpha_{0}\right)  $ to the same order in perturbations
should be taken to be
\begin{equation}
N\left(  \alpha_{0}\right)  =\alpha_{0}+3\alpha_{0}^{2}+24\alpha_{0}%
^{3}+285\alpha_{0}^{4}+4284\alpha_{0}^{5}+75\,978\alpha_{0}^{6}+O\left(
\alpha_{0}^{7}\right)  . \label{13a}%
\end{equation}
The\textit{ effective running coupling} as a function of distance can be
written then as perturbative series in powers of $\alpha_{0}$
\begin{equation}
\alpha_{eff}\left(  x\right)  =\alpha_{eff}\left(  r\right)  =\alpha_{0}%
f^{2}\left(  r\right)  =\sum\limits_{n=0}^{\infty}\alpha_{0}^{n+1}g_{n}\left(
x\right)  , \label{14a}%
\end{equation}
where
\[
g_{0}\left(  x\right)  =1,\text{ }g_{1}\left(  x\right)  =2x,\text{ }%
g_{2}\left(  x\right)  =4x^{2}+6x,\text{ }g_{3}\left(  x\right)
=8x^{3}+30x^{2}+48x,
\]%
\[
g_{4}\left(  x\right)  =16x^{4}+104x^{3}+342x^{2}+570x,
\]%
\[
g_{5}\left(  x\right)  =32x^{5}+308x^{4}+1572x^{3}+4998x^{2}+8568x,
\]%
\begin{equation}
g_{6}\left(  x\right)  =64x^{6}+\frac{4176}{5}x^{5}+5880x^{4}+27\,612x^{3}%
+86\,832x^{2}\allowbreak+151\,956x, \label{18a}%
\end{equation}
etc. The calculation of $g_{n}\left(  x\right)  $ is straightforward and we
did it until $g_{10}\left(  x\right)  .$ However, to simplify the formulae we
present here the result only up to $g_{6}\left(  x\right)  .$ Note that the
running coupling constant depends on $r$ only logarithmically with the
coefficients $g_{n}$ power series of $x=\ln\left(  r/r_{0}\right)  $ with the
highest power $n.$ Moreover, the series $\left(  \ref{14a}\right)  $ can be
rearranged and partially resummed. In particular, collecting together leading
powers of logarithms, next-to-leading and next-to-next leading powers we get%
\begin{align}
\alpha_{eff}\left(  x\right)   &  =\alpha_{0}\left[  1+2\tilde{x}+4\tilde
{x}^{2}+8\tilde{x}^{4}+16\tilde{x}^{5}+32\tilde{x}^{6}+O\left(  \tilde{x}%
^{7}\right)  \right] \nonumber\\
&  +\alpha_{0}^{2}\left[  6\tilde{x}+30\tilde{x}^{2}+104\tilde{x}%
^{3}+308\tilde{x}^{4}+\frac{4176}{5}\tilde{x}^{5}+\frac{10\,704}{5}%
\allowbreak\tilde{x}^{6}+O\left(  \tilde{x}^{7}\right)  \right] \nonumber\\
&  +\alpha_{0}^{3}\left[  \allowbreak48\tilde{x}+342\tilde{x}^{2}%
+1572\tilde{x}^{3}+5880\tilde{x}^{4}+\frac{97\,248}{5}\tilde{x}^{5}%
+\allowbreak59\,248\tilde{x}^{6}+O\left(  \tilde{x}^{7}\right)  \right]
\nonumber\\
&  +O\left(  \alpha_{0}^{4}...\right)  \label{20a}%
\end{align}
where $\tilde{x}=\alpha_{0}x.$ In the second and third brackets we have also
included higher order terms compared to $\left(  \ref{18a}\right)  .$ The
series in the bracket can be resummed. In particular, it is obvious that%
\begin{equation}
1+2\tilde{x}+4\tilde{x}^{2}+8\tilde{x}^{4}+16\tilde{x}^{5}+32\tilde{x}%
^{6}+...=\frac{1}{1-2\tilde{x}}, \label{21a}%
\end{equation}
Much less obvious are the following results%
\begin{equation}
6\tilde{x}+30\tilde{x}^{2}+104\tilde{x}^{3}+308\tilde{x}^{4}+\frac{4176}%
{5}\tilde{x}^{5}+...=\frac{3\ln\left(  1-2\tilde{x}\right)  }{\left(
1-2\tilde{x}\right)  ^{2}}, \label{22a}%
\end{equation}
and%
\begin{align}
&  48\tilde{x}+342\tilde{x}^{2}+1572\tilde{x}^{3}+5880\tilde{x}^{4}%
+\frac{97\,248}{5}\tilde{x}^{5}+\allowbreak59\,248\tilde{x}^{6}...\nonumber\\
&  =\frac{9\left(  \ln\left(  1-2\tilde{x}\right)  \right)  ^{2}-9\ln\left(
1-2\tilde{x}\right)  +30\tilde{x}}{\left(  1-2\tilde{x}\right)  ^{3}},
\label{24a}%
\end{align}
which the reader can verify just expanding the appropriate expressions in
powers of $\tilde{x}.$ One may wonder how did we manage to resum these last
two series? The answer to this question is in the next section where we
uncover the renormalization group structure of our entirely classical theory
and derive the $\beta$ function which generates the resummation of the
perturbative expansion to the appropriate powers of $\alpha_{0}.$

\section{Renormalization group and asymptotic freedom}
In the Wilsonian approach \cite{wilson} the renormalization group sets how the
couplings of the quantum theory should change under re-scalings of the UV
cutoff . The equations governing this dependence are known as the
renormalization group equations. This general notion of renormalization group
can be extended to the classical field theory with external point-like sources
in the following sense. Let us introduce an UV cutoff $r_{0}$ setting the way
we smear the source. The classical field created by such source will
generically depends on the regulator $r_{0}$ and the self-coupling $\lambda$
of the theory. One can ask under which circumstances we can require that the
classical theory must be invariant under re-scaling of $r_{0}$. This is
possible only if the corresponding classical theory incorporates the
renormalization group structure. In this case the dependence of the coupling
on the \textit{smearing cutoff} $\ \lambda(r_{0})$ also captures the screening
and anti-screening effects. Moreover, using the effective running coupling we
can associate with an external source, a physical length scale $R_{c}$ by the
standard procedure of dimensional transmutation.

The reason why the classical solution captures the renormalization group
structure is easy to understand. Any regularization scheme in quantum field
theory give rise to logarithmic contributions which even in a scale invariant
theory lead to anomalous scaling. These logarithmic contributions are of the
type $\log\left(  p/\Lambda\right)  $ with $\Lambda$ the UV cutoff. \ Since
the divergent contribution $\log\Lambda$ is absorbed by renormalization, we
are free to choose the scale $p$ at which the logarithmic contribution to the
self-energy vanishes. As a consequence the scaling of $\Lambda$ should be
accompanied by finite renormalizations (the RG transformations) of the
coupling constants. In the classical theory under consideration we have found
the same type of logarithmic contributions to the field created by the
external source. In this case the role of cutoff $\Lambda$ is played by the
smearing scale $r_{0}$. One can renormalize the classical theory by
subtracting the $\log r_{0}$ contributions as it is done in quantum field
theory. However, if we want physics to be independent on the method of
removing this infinity, we need to change the couplings, exactly as it is done
in quantum field theory. Both renormalization group structures, the classical
and the quantum ones, are structurally identical for $\lambda\phi^{4}$ theory
because both have the same type of parent logarithmic contributions.

\subsection{Perturbative $\beta$ function}
Once we have obtained the classical expression $\left(  \ref{14a}\right)  $
for the effective coupling constant $\alpha_{eff}\left(  r\right)  $ we can
check whether taking the bare coupling constant $\alpha_{0}$ as a function of
$r_{0}$ we can make $\alpha_{eff}\left(  r\right)  $ independent of $r_{0}.$
As we have said, this is possible only in the theories with associated
renormalization group structure, which in turn imposes rather severe
conditions on the functions $g_{n}\left(  x\right)  $ in the perturbative
expansion $\left(  \ref{14a}\right)  .$ Let us first derive these conditions,
which do not depend on the origin (classical or quantum) of the
renormalization group, and then verify whether they are satisfied by the
functions in $\left(  \ref{18a}\right)  $.

On general grounds the expansion of the dimensionless running coupling
constant $\alpha_{eff}\left(  r\right)  $ in powers of $\alpha_{0}%
=\alpha\left(  r_{0}\right)  ,$ normalized at $r=r_{0},$ can be written as%
\begin{equation}
\alpha_{eff}\left(  r\right)  =\alpha\left(  r_{0}\right)  +\alpha^{2}\left(
r_{0}\right)  g_{1}\left(  \frac{r}{r_{0}}\right)  +...=\sum\limits_{n=0}%
^{\infty}\alpha^{n+1}\left(  r_{0}\right)  g_{n}\left(  \frac{r}{r_{0}%
}\right)  , \label{23a}%
\end{equation}
where we use the spatial scale $r$ instead of the usually used energy scale
$k\sim1/r$. It is clear that $g_{0}\left(  r/r_{0}\right)  =1$ and since
$\alpha\left(  r\right)  =\alpha\left(  r_{0}\right)  $ at $r=r_{0},$ we have%
\begin{equation}
g_{n}\left(  1\right)  =0, \label{24}%
\end{equation}
for $n\geq1.$ Invariance under changes of the ultraviolet regulator $r_{0},$
implies%
\begin{equation}
\frac{d}{dr_{0}}\left(  \sum\limits_{n=0}^{\infty}\alpha^{n+1}\left(
r_{0}\right)  g_{n}\left(  \frac{r}{r_{0}}\right)  \right)  =0, \label{25}%
\end{equation}
which in turn imposes severe restrictions on $g_{n}\left(  r/r_{0}\right)  .$
Taking the derivative and rearranging the terms in $\left(  \ref{25}\right)  $
leads to%
\begin{equation}
\frac{d\alpha\left(  r_{0}\right)  }{d\ln r_{0}}=\alpha^{2}\left(
r_{0}\right)  \frac{\sum\limits_{k=0}^{\infty}g_{k+1}^{\prime}\left(
x\right)  \alpha^{k}\left(  r_{0}\right)  }{\sum\limits_{k=0}^{\infty}\left(
k+1\right)  g_{k}\left(  x\right)  \alpha^{k}\left(  r_{0}\right)  },
\label{26}%
\end{equation}
where $x=\ln\left(  r/r_{0}\right)  $ and prime denotes the derivative with
respect to $x.$ The ratio of sums in the right hand side of $\left(
\ref{26}\right)  $ should not depend on $x$ because the left hand side of this
equation is $x$-independent. Therefore setting $x=0$ (which corresponds to
$r=r_{0}$) and taking into account $\left(  \ref{24}\right)  $ we find that
\begin{equation}
\frac{\sum\limits_{k=0}^{\infty}g_{k+1}^{\prime}\left(  x\right)  \alpha
^{k}\left(  r_{0}\right)  }{\sum\limits_{k=0}^{\infty}\left(  k+1\right)
g_{k}\left(  x\right)  \alpha^{k}\left(  r_{0}\right)  }=\sum\limits_{k=0}%
^{\infty}g_{k+1}^{\prime}\left(  0\right)  \alpha^{k}\left(  r_{0}\right)  ,
\label{27}%
\end{equation}
from where it follows that the functions $g_{k}\left(  x\right)  $ should
satisfy the following recursion relations:%
\begin{equation}
\frac{dg_{n+1}\left(  x\right)  }{dx}=\sum\limits_{k=0}^{n}\left(  k+1\right)
g_{n+1-k}^{\prime}\left(  0\right)  g_{k}\left(  x\right)  . \label{28}%
\end{equation}
Note that only if these conditions are satisfied then there exists a function
$\alpha\left(  r_{0}\right)  $ for which the sum in the right hand side of
$\left(  \ref{23a}\right)  $ does not depend on $r_{0}.$ Nicely enough the
unambiguous solution of these recursion relations with \textquotedblleft
initial conditions\textquotedblright\ (\ref{24}) is given by%
\begin{equation}
g_{n}\left(  x\right)  =\sum\limits_{k=1}^{n}c_{k}x^{k}, \label{29}%
\end{equation}
where $c_{k}$ are completely determined by the numerical values of
$g_{1}^{\prime}\left(  0\right)  ,g_{2}^{\prime}\left(  0\right)  ,..$ which
in principle can be arbitrary. For instance, the coefficient in front of the
leading logarithm $x^{n}=\ln^{n}\left(  r/r_{0}\right)  $ in $g_{n}$ is equal
to $c_{n}=\left(  g_{1}^{\prime}\left(  0\right)  \right)  ^{n}.$

At this point it is quite rewarding to confirm that the set of classical
functions $\left(  \ref{18a}\right)  $ in fact satisfies the recursion
relations $\left(  \ref{28}\right)  .$ This can be done by direct calculation
to any order in perturbation theory (we did it up to $g_{10}$). Thus, taking
$\alpha_{0}$ in $\left(  \ref{14a}\right)  $ to be the function of $r_{0}$ we
uncover the renormalization group structure of the classical $\lambda\phi^{4}$
theory. We would like to stress that in distinction from the quantum field
theory, where the renormalization group is normally checked by direct
calculations only to the leading logarithms (and postulated otherwise), we
verified it also for all subleading logarithms.

To take the advantage of renormalization group for partial resummation of the
perturbative expansion $\left(  \ref{14a}\right)  $ we note that from $\left(
\ref{26}\right)  $ and $\left(  \ref{27}\right)  $ it follows
\begin{equation}
\frac{d\alpha\left(  r_{0}\right)  }{d\ln r_{0}}=\alpha^{2}\left(
r_{0}\right)  \sum\limits_{k=0}^{\infty}g_{k+1}^{\prime}\left(  0\right)
\alpha^{k}\left(  r_{0}\right)  . \label{30}%
\end{equation}
The running constant $\alpha_{eff}\left(  r\right)  $ depends on $r$ in the
same way that $\alpha\left(  r_{0}\right)  $ depends on $r_{0}.$ Hence
$\alpha_{eff}\left(  x\right)  $ satisfies the well known Gell-Mann-Low
equation \cite{GML}%
\begin{equation}
\frac{d\alpha_{eff}\left(  x\right)  }{dx}=\alpha_{eff}^{2}\left(  x\right)
\sum\limits_{k=0}^{\infty}g_{k+1}^{\prime}\left(  0\right)  \alpha_{eff}%
^{k}\left(  x\right)  . \label{31}%
\end{equation}
The $\beta$ function is normally defined as the derivative of $\alpha_{eff}$
with respect to the logarithm of the energy squared. For us it is more
convenient to define it as%
\begin{equation}
\beta\equiv\frac{d\alpha_{eff}\left(  x\right)  }{dx}, \label{32}%
\end{equation}
which (up to factors $4\pi$ due to the choice of charge units) is related to
the standard $\beta_{st}$ function as $\beta_{st}=-\beta/2.$ According to
$\left(  \ref{31}\right)  $ and $\left(  \ref{18a}\right)  $ the classical
perturbative $\beta$ function is equal to%
\begin{align}
\beta\left(  \alpha\right)   &  =\sum\limits_{k=1}^{\infty}\beta_{k}%
\alpha^{k+1}=2\alpha^{2}+6\alpha^{3}+48\alpha^{4}+570\alpha^{5}\nonumber\\
&  +8568\alpha^{6}+151956\alpha^{7}+..., \label{33a}%
\end{align}
where $\alpha\equiv\alpha_{eff}\left(  x\right)  $ and $\beta_{i}\equiv
g_{i}^{\prime}\left(  0\right)  .$

Obviously, we should not expect the numerical coefficients $\beta_{i}$ of this
\textit{classical beta function} to be identical to the ones derived in the
quantum field theory. In the last case $\beta_{i}$ are determined by the loop
contributions and they will depend, beyond two loops, on the particular
renormalization scheme used to segregate a finite part of the divergent loop
integrals. The \textit{classical} beta function accounts for the
anti-screening effects due to the classical self-interaction. A potential
quantum theory check of the numerical coefficients derived above will require
to work in the presence of large external charge where we have to modify the
Green functions in order to account for the effect of the external charge.
Because for $Q\gg1$ the quantum fluctuations are subdominant we expect that
classical contribution dominates. Although the direct check of this
expectation is obviously important we will not follow that path. Instead we
will restrict ourselves to the physical consequences of the underlying
renormalization group structure of the classical theory.

\subsection{Partial resummations}
In equation $\left(  \ref{20a}\right)  $ we have separately collected the
contribution of the leading and subleading logarithms to $\alpha\left(
x\right)  $ and presented the result of their resummation. For the subleading
logarithm the result was derived using Gell-Mann-Low equation. For $\alpha
\ll1$ we can first neglect all terms in $\beta$ function besides of the
\textquotedblleft one loop\textquotedblright\ contribution. Equation $\left(
\ref{31}\right)  $ then reduces to%
\begin{equation}
\frac{d\alpha\left(  x\right)  }{dx}=2\alpha^{2}\left(  x\right)  , \label{34}%
\end{equation}
and its solution, with initial condition $\alpha\left(  0\right)  =\alpha
_{0}\equiv-\lambda_{0}Q^{2},$ is
\begin{equation}
\alpha\left(  r\right)  =\frac{\alpha_{0}}{1-2\alpha_{0}x}=\frac{-\lambda
_{0}Q^{2}}{1+2\lambda_{0}Q^{2}\ln\left(  r/r_{0}\right)  }, \label{35}%
\end{equation}
where $\lambda_{0}=\lambda\left(  r_{0}\right)  <0.$ It is easy to see that
this solution gives us the resummation of the leading logarithms in the
expansion $\left(  \ref{20a}\right)  ,\left(  \ref{21a}\right)  .$

We can repeat the same analysis keeping in Gell-Mann-Low equation the
contribution up to two loops,%
\begin{equation}
\frac{d\alpha\left(  x\right)  }{dx}=2\alpha^{2}\left(  x\right)  +6\alpha
^{3}. \label{38}%
\end{equation}
Integrating this equation with initial condition $\alpha\left(  0\right)
=\alpha_{0}$, we obtain%
\begin{equation}
\frac{1}{\alpha\left(  x\right)  }-3\ln\left(  \frac{1+3\alpha\left(
x\right)  }{1+3\alpha_{0}}\times\frac{\alpha_{0}}{\alpha\left(  x\right)
}\right)  =\frac{1-2\alpha_{0}x}{\alpha_{0}}. \label{39}%
\end{equation}
Solving this equation in terms of the perturbative expansion in $\alpha_{0}$
one gets%
\begin{equation}
\alpha\left(  x\right)  =\frac{\alpha_{0}}{1-2\alpha_{0}x}-3\left(
\frac{\alpha_{0}}{1-2\alpha_{0}x}\right)  ^{2}\ln\left(  1-2\alpha
_{0}x\right)  +O\left(  \alpha_{0}^{3}\right)  \label{40}%
\end{equation}
Note that the second term agrees with resummation $\left(  \ref{22a}\right)  $
of the next to the leading order logarithms$.$ The same is true at three loop
order, where the solution to
\begin{equation}
\frac{d\alpha\left(  x\right)  }{dx}=2\alpha^{2}\left(  x\right)  +6\alpha
^{3}+48\alpha^{4}, \label{40a}%
\end{equation}
which is%
\begin{align}
\alpha\left(  x\right)   &  =\frac{\alpha_{0}}{1-2\alpha_{0}x}-3\left(
\frac{\alpha_{0}}{1-2\alpha_{0}x}\right)  ^{2}\ln\left(  1-2\alpha_{0}x\right)
\label{42a}\\
&  +9\left(  \frac{\alpha_{0}}{1-2\alpha_{0}x}\right)  ^{3}\left(  \ln
^{2}\left(  1-2\alpha_{0}x\right)  -\ln\left(  1-2\alpha_{0}x\right)
+\frac{30}{9}\alpha_{0}x\right)  +O\left(  \alpha_{0}^{4}\right)  ,\nonumber
\end{align}
also accounts for the resummation of next-to-next subleading logarithms. In
other words the solutions to the classical renormalization group equation give
us the resummation of the perturbative series $\left(  \ref{14a}\right)  $
taking care in every step about next logarithms in $g_{n}\left(  x\right)  .$
It is clear that when the running coupling constant becomes of order unity
(strong coupling regime) all terms in expansion $\left(  \ref{42a}\right)  $
are of the same order and the series $\left(  \ref{42a}\right)  $ should be
further resummed. It is not a priori clear whether the singularity in this
expansion (Landau pole\cite{landau}) will survive after this resummation. We
will answer this question in the next section using nonperturbative methods.

\subsection{Dimensional transmutation and asymptotic freedom}
One important consequence of the renormalization group is dimensional
transmutation. We can easily understand this phenomenon \ using the result of
the one loop resummation of perturbative expansion%
\begin{equation}
\alpha\left(  r\right)  =\frac{-\lambda_{0}Q^{2}}{1+2\lambda_{0}Q^{2}%
\ln\left(  r/r_{0}\right)  }. \label{42b}%
\end{equation}
In this expression $\lambda_{0}$ depends on the regulator $r_{0}$ in such a
way that $\alpha\left(  r\right)  $ is $r_{0}$-independent to the
corresponding order. Therefore we can define the renormalization group
invariant physical scale $R_{c}$ via%
\begin{equation}
\ln\frac{R_{c}}{r_{0}}=-\frac{1}{2\lambda_{0}Q^{2}}. \label{42c}%
\end{equation}
Note that this scale%
\begin{equation}
R_{c}=r_{0}e^{-\frac{1}{2\lambda\left(  r_{0}\right)  Q^{2}}}, \label{42d}%
\end{equation}
does not depend on the particular value of regulator $r_{0}$ at one loop
level. Using this \textit{dynamically generated scale }we can rewrite the
physical running coupling as%
\begin{equation}
\alpha\left(  r\right)  \equiv-\lambda\left(  r\right)  Q^{2}=\frac{1}%
{2\ln\left(  R_{c}/r\right)  }.
\end{equation}
The physical meaning of this expression is obvious. Perturbatively the theory
can be defined in the ultraviolet region corresponding to length scales $r\ll
R_{c},$ where it becomes effectively free. Thus, we have found
\textit{asymptotic freedom} in the \textit{classical} $\lambda\phi^{4}$ theory
with negative $\lambda.$

In the infrared at length scales of order $R_{c}$ the theory becomes strongly
coupled and non-perturbative. What is the potential meaning of this
dynamically generated scale? From the point of view of the classical theory
the existence of this scale is quite surprising since it is independent of the
UV regulator. On the top of that $R_{c}$ is a very non-perturbative scale.
Obviously it is tempting to think of $R_{c}$ as setting the natural
confinement scale of the theory. A way to check this claim is to derive an
exact \textit{classical} $\beta$ function equation and to read off the
previous perturbative expansion from the corresponding solution of this
equation. We address these issues below.

\section{Beyond perturbation theory and asymptotic behavior}
The usual way to address the non-perturbative phenomena within perturbation
theory is to study the convergence of the perturbative series. In
\cite{lipatov} it was found that the numerical coefficients in the
perturbative expansion of $\beta$ function asymptotically grow as $\beta
_{k}\sim k!\beta_{1}^{k},$ where $\beta_{1}$ is one loop $\beta$ function and
$k$ denotes the perturbative order in coupling constant. Such behavior sets
the limit of perturbation theory and fixes the uncertainty of the computations
to be of order $\exp\left(  -1/\beta_{1}\alpha\right)  .$ In the theories with
asymptotic freedom this uncertainty is extremely small in the deep UV region
contrary to what happens in the theories with UV Landau pole. Normally in
quantum field theory is hard to prove this factorial asymptotic behavior of
the coefficients in $\beta$ function. It can have different origin: either the
growth of the number of diagrams contributing to a given order in perturbation
theory (instanton effect) or the contribution of multi-bubble diagrams (renormalons).

In the classical approach to the renormalization group, however, there is an
opportunity to convert the classical equations of motion into exact equation
for the $\beta$ function. This equation be can used afterwards to check the
similarity between classical and quantum renormalization groups. In
particular, as we will see, one can use the exact equation to derive the
asymptotic behavior of the coefficients in the perturbative expansion of
$\beta$ function. Interestingly enough the asymptotic behavior anticipated by
the exact classical $\beta$ function agrees with the quantum filed theory
expectations. In addition this allows us to clarify the origin of renormalons
as well as the generic form of the non-perturbative uncertainties. Moreover,
the non-perturbative contributions will be naturally defined in terms of the
dynamically generated scale $R_{c},$ as it should be.

\subsection{Exact classical $\beta$ function}
To derive an exact equation for the classical $\beta$ function we begin with
equation for static scalar field outside an external source%
\begin{equation}
\frac{1}{r^{2}}\frac{d}{dr}\left(  r^{2}\frac{d\phi}{dr}\right)  -\lambda
_{0}\phi^{3}=0. \label{43}%
\end{equation}
Substituting
\begin{equation}
\phi=\frac{Qf\left(  r\right)  }{r}, \label{44}%
\end{equation}
we can rewrite the equation above as%
\begin{equation}
f^{\prime\prime}-f^{\prime}+\alpha_{0}f^{3}=0, \label{45}%
\end{equation}
where $\alpha_{0}=-\lambda_{0}Q^{2}$ and prime denotes the derivative with
respect to $x=\ln\left(  r/R_{c}\right)  .$ Multiplying this equation by
$\alpha_{0}f$ and defining running coupling constant as before%
\begin{equation}
\alpha\left(  x\right)  =\alpha_{0}f^{2}\left(  x\right)  , \label{46}%
\end{equation}
we obtain the following second order differential equation for $\alpha\left(
x\right)  $%
\begin{equation}
\alpha^{\prime\prime}-\frac{\alpha^{\prime2}}{2\alpha}-\alpha^{\prime}%
+2\alpha^{2}=0. \label{47}%
\end{equation}
Recalling the definition of $\beta$ function, $\beta\equiv\alpha^{\prime},$
and taking into account that%
\begin{equation}
\alpha^{\prime\prime}=\alpha^{\prime}\frac{d\alpha^{\prime}}{d\alpha}%
=\beta\frac{d\beta}{d\alpha}, \label{48}%
\end{equation}
equation $\left(  \ref{47}\right)  $ reduces to the first order differential
equation:%
\begin{equation}
\beta=2\alpha^{2}+\frac{1}{2}\left(  \frac{d\beta^{2}}{d\alpha}-\frac
{\beta^{2}}{\alpha}\right)  , \label{49}%
\end{equation}
which determines the exact classical $\beta$ function.

\subsection{Weak coupling expansion and renormalons}
Let us first use the exact equation for $\beta$ function to reproduce our
perturbative results above. In order to do that we substitute in $\left(
\ref{49}\right)  $%
\begin{equation}
\beta\left(  \alpha\right)  =\sum\limits_{k=1}^{\infty}\beta_{k}\alpha^{k+1}.
\label{49a}%
\end{equation}
This leads to the following recursion relations for the unknown numerical
coefficients $\beta_{k}$%
\begin{equation}
\beta_{1}=2,\text{ \ \ }\beta_{k}=\sum\limits_{m=1}^{k-1}\left(  m+\frac{1}%
{2}\right)  \beta_{k-m}\beta_{m}\text{ for }k\geq2, \label{49b}%
\end{equation}
Using this relations we find
\begin{equation}
\beta_{1}=2,\text{ }\beta_{2}=6,\text{ }\beta_{3}=48,\text{ }\beta
_{4}=570,\text{ }\beta_{5}=8568,\text{ }\beta_{6}=151956,... \label{49c}%
\end{equation}
in complete agreement with $\left(  \ref{33a}\right)  $ to an arbitrary order
in $\alpha.$ Thus we have proven that equation $\left(  \ref{49}\right)  $
yields the exact $\beta$ function which is in complete agreement with the
perturbative $\beta$ function.

Let us now find the asymptotic behavior of the perturbative series. One can
easily see that for large $k$ the main contribution to the sum in $\left(
\ref{49b}\right)  $ comes from the terms with $m=k-1$ and $m=1$ and the
recursion relation reduces to%
\begin{equation}
\beta_{k}\simeq\left(  k+1\right)  \beta_{1}\beta_{k-1}, \label{49d}%
\end{equation}
the solution of which is%
\begin{equation}
\beta_{k}\simeq\left(  k+1\right)  !\beta_{1}^{k} \label{49e}%
\end{equation}
with $\beta_{1}=2.$ Nicely enough this is the same type of factorial behavior
we expect in quantum field theory. Moreover, since the coefficient of the
factorial is the one-loop $\beta$ function it is natural to identify the
origin of this behavior with a renormalon.

In order to clarify the meaning of this renormalon let us consider the
perturbative solution of equation $\left(  \ref{49}\right)  $ assuming that
$\alpha\ll1.$ Substituting%
\begin{equation}
\beta=2\alpha^{2}\left(  1+\varepsilon\right)  , \label{51a}%
\end{equation}
in $\left(  \ref{49}\right)  $ we find that $\varepsilon\left(  \alpha\right)
$ satisfies the equation%
\begin{equation}
2\alpha^{2}\frac{d\varepsilon}{d\alpha}=\frac{\varepsilon}{1+\varepsilon
}-3\alpha\left(  1+\varepsilon\right)  =\varepsilon-3\alpha+O\left(
\varepsilon^{2},\varepsilon\alpha\right)  . \label{51b}%
\end{equation}
Because both $\varepsilon$ and $\alpha$ are much less than unity we can
neglect nonlinear terms. Solving the resulting linear equation one obtains%
\begin{equation}
\varepsilon\left(  \alpha\right)  =\left(  \frac{3}{\beta_{1}}%
\operatorname{Ei}\left(  \frac{1}{\beta_{1}\alpha}\right)  +C\right)
e^{-\frac{1}{\beta_{1}\alpha}}+O\left(  \left(  e^{-\frac{1}{\beta_{1}\alpha}%
}\right)  ^{2}\right)  , \label{51c}%
\end{equation}
where $\beta_{1}=2,$ $\operatorname{Ei}\left(  z\right)  $ is the
exponential-integral function and $C$ is the constant of integration. Now
using the asymptotic expansion of the exponential-integral function at large
argument%
\begin{equation}
\operatorname{Ei}\left(  z\right)  =\frac{e^{z}}{z}\left(  \sum\limits_{k=0}%
^{n}\frac{k!}{z^{k}}+O\left(  \frac{1}{z^{n+1}}\right)  \right)  , \label{51d}%
\end{equation}
we find that asymptotically the coefficients of the $\beta$ function in the
perturbative expansion grow as $\beta_{k}\sim k!\beta_{1}^{k}.$ This
completely clarifies the origin of the renormalon, which is an
\textit{artifact} of the asymptotic expansion of non-analytic function. It
also follows from $\left(  \ref{51c}\right)  $ that the accuracy of the Borel
resummation does not exceed
\begin{equation}
e^{-\frac{1}{\beta_{1}\alpha}}\sim\frac{r}{R_{c}}, \label{52a}%
\end{equation}
which is the expected non-perturbative uncertainty!

\section{The nonperturbative solution and confinement}
Finally let us use the exact $\beta$ function equation to determine what
happens in the infrared region $r>R_{c},$ where the perturbation theory is
completely out of the control. Obviously the nonperturbative effective
coupling constant can be used to define a static \textit{inter-quark}
potential. Regarding the nature of the classical sources we will leave beyond
the scope of paper and concentrate mostly on the calculation of the behavior
of the running coupling constant in the infrared region. As we will see this
coupling constant determines-similar to QCD- the confining potential between
sources. In particular we will compute the energy of an isolated source and a
dipole, built out of two opposite charges. In the first case we find the
divergent energy which is an indication of confinement and can be interpreted
as the absence of isolated sources. The energy of the dipole is finite and
positive. Its typical size is of order $R_{c}$ and the binding energy $\sim
R_{c}^{-1}$ in case of sources with negligible masses. This is an encouraging
hint towards explaining the colorless hadron states in QCD.

\subsection{The infrared coupling constant \ }
\begin{figure}[hbt]
\centering
\psfrag{x}{$\alpha$}\psfrag{y}{$\beta(\alpha)$}\psfrag{x1}[tc][cc]{$10^{-2}$}\psfrag{x2}[tc][cc]{$10^{-1.5}$}\psfrag{x3}[tc][cc]{$10^{-1}$}\psfrag{xmax}[tc][cc]{$0.69$}\psfrag{xend}[tc][cc]{$0.98$}
\includegraphics[width=0.8\textwidth]{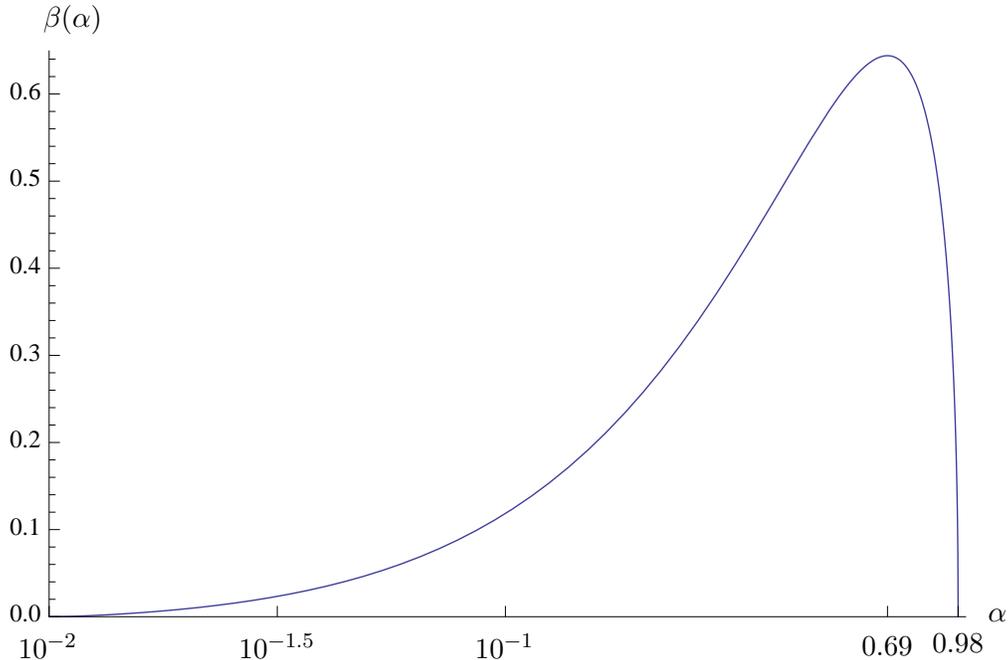}
\caption{Numerical evaluation of nonperturbative $\beta$ function}
\label{fig:Fig1}
\end{figure}

Equation $\left(  \ref{49}\right)  $ for the exact $\beta$ function can
rewritten in the form
\begin{equation}
\frac{d\beta}{d\alpha}=\frac{2\alpha\beta-4\alpha^{3}+\beta^{2}}{2\alpha\beta
}, \label{52}%
\end{equation}
and can be fully investigated using the phase diagram method. The particular
solution we need is determined by the perturbative initial condition $\left(
\ref{33a}\right)  .$ The resulting nonperturbative $\beta$ function, shown in
Fig.\ \ref{fig:Fig1}, reaches a maximum value about $0.64$ for $\alpha\simeq$ $0.69$ and
after that decreases and vanishes at $\alpha\approx0.98.$ 
Such behavior of the
$\beta$ function in nonperturbative regime is quite nontrivial. The most
dramatic effect is the absence of any divergence for the coupling constant,
meaning that in case of exact $\beta$ function, which accounts for all
resummations, the IR Landau pole is absent. This nonperturbative resummations
leads to finite running coupling for any finite $x$ interpolating smoothly
between the asymptotically free UV-regime and the IR- region $r\gg R_{c}$.

\begin{figure}[hbt]
\centering
\psfrag{x}{$r$}\psfrag{y}{$\alpha(r)$}\psfrag{1}[cr][cr]{1}\psfrag{Rc}[tc][cc]{$R_{c}$}\psfrag{pert}{\begin{tabular}{l}perturbative\\ result\end{tabular}}\psfrag{prop}{$\propto r^{2/3}$}
\includegraphics[width=0.8\textwidth]{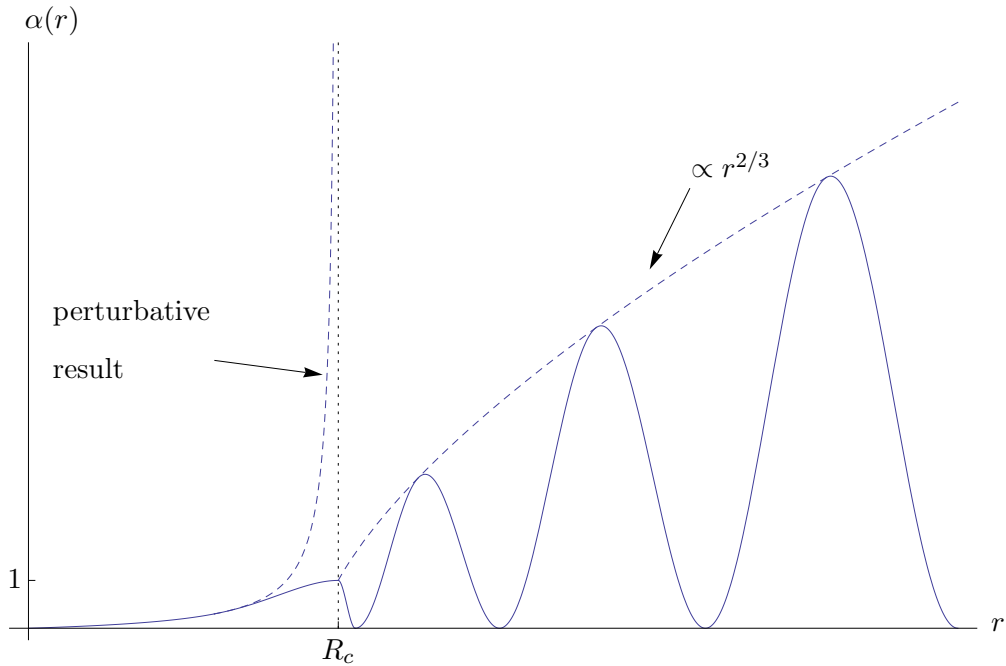}
\caption{Nonperturbative running coupling}
\label{fig:Fig2}
\end{figure}

Beyond the first zero of the $\beta$ function it is more convenient to draw
$\alpha$ and $\beta$ separately as functions of $x=\ln\left(  r/R_{c}\right)
$ because both $\alpha$ and $\beta$ become oscillating functions of the scale.
In order to find the solution in this region it is more convenient to work
directly with equation $\left(  \ref{45}\right)  $ instead of $\left(
\ref{52}\right)  .$ This is the equation for a particle \textquotedblleft
moving\textquotedblright\ in a positive quartic potential in the presence of
negative friction. Such particle \textquotedblleft
oscillates\textquotedblright\ and if we neglect for a moment the friction
term, the typical \textquotedblleft period of oscillation\textquotedblright%
\ can be estimated on dimensional grounds as
\[
\Delta x\sim\sqrt{\frac{1}{\alpha_{0}f^{2}}}\sim\sqrt{\frac{1}{\alpha\left(
x\right)  }},
\]
where $f$ is the typical amplitude of oscillations. It is clear that this
estimate is valid only if $\Delta x\ll1$ because otherwise friction term
dominates and completely damps the oscillations. However, for $\alpha\left(
x\right)  \gg1,$ when $\Delta x\ll1,$ the friction is not so crucial and the
system undergoes oscillations with the amplitude slowly growing due to this
negative friction. In order to find how fast this amplitude grows we multiply
equation $\left(  \ref{45}\right)  $ by $f$ and rewrite it in the form%
\begin{equation}
\left(  ff^{\prime}\right)  ^{\prime}-f^{\prime2}-\frac{1}{2}\left(
f^{2}\right)  ^{\prime}+\alpha_{0}f^{4}=0. \label{53}%
\end{equation}
Averaging this equation over the period of oscillations we find%
\begin{equation}
\left\langle f^{\prime2}\right\rangle =\alpha_{0}\left\langle f^{4}%
\right\rangle . \label{54}%
\end{equation}
Multiplying now $\left(  \ref{45}\right)  $ by $f^{\prime}$ leads to the
equation%
\begin{equation}
\left(  \frac{1}{2}f^{\prime2}+\frac{1}{4}\alpha_{0}f^{4}\right)  ^{\prime
}=f^{\prime2}, \label{55}%
\end{equation}
which after averaging and taking into account $\left(  \ref{54}\right)  $
\ gives us%
\begin{equation}
\frac{d\left\langle f^{4}\right\rangle }{dx}=\frac{4}{3}\left\langle
f^{4}\right\rangle . \label{56}%
\end{equation}
Finally solving this equation we obtain
\begin{equation}
\left\langle f^{4}\right\rangle =C\exp\left(  \frac{4x}{3}\right)  =C\left(
\frac{r}{R_{c}}\right)  ^{4/3}, \label{57}%
\end{equation}
leading to the following nonperturbative behavior of the running coupling
constant
\begin{align}
\alpha\left(  r\right)   &  =\alpha_{0}f^{2}\simeq\alpha_{0}\sqrt{\left\langle
f^{4}\right\rangle }\cos^{2}\left(  \int\sqrt[4]{\alpha_{0}^{2}\left\langle
f^{4}\right\rangle }dx\right) \nonumber\\
&  \simeq O\left(  1\right)  \left(  \frac{r}{R_{c}}\right)  ^{2/3}\cos
^{2}\left(  \frac{r}{R_{c}}\right)  ^{2/3}, \label{59}%
\end{align}
for $r\gg R_{c}.$ In Fig.\ \ref{fig:Fig2}  we summarize the behavior of the running coupling.
As it was said above although according to perturbation theory the coupling
constant should become infinite at $r=R_{c},$ in reality this does not happen.
The coupling constant remains finite and at scales larger than the confinement
scale grows as $r^{2/3}.$ This is the main non-perturbative result we can
extract from the exact $\beta$ function equation.

\subsection{Confinement}
As already discussed we can mimic confinement identifying the classical
sources as \textit{static quarks} and defining a quench approximation to the
static \textit{inter-quark} potential in terms of the effective coupling in
the IR region.

Since in our case the self-interaction contribution to the energy goes like
$\phi^{4}\sim f^{4}\sim\alpha(r)^{2}$ we can define the static potential as:
\begin{equation}
V(r)\sim\frac{\alpha^{2}(r)}{r} \label{59b}%
\end{equation}
Using the non-perturbative value of the running coupling (\ref{59}) we get
\begin{equation}
V(r)\sim O\left(  1\right)  R_{c}^{-1}\left(  \frac{r}{R_{c}}\right)  ^{1/3}
\label{59c}%
\end{equation}
To check this qualitative result let us compute the energy of the field
created by a static external charge. Since the field is static and spherically
symmetric, the total energy is given by the expression%

\begin{align}
E  &  =\frac{1}{2}\int\left(  \left(  \nabla\phi\right)  ^{2}+\frac
{\lambda_{0}}{2}\phi^{4}\right)  d^{3}x=2\pi\int\left(  \left(  \partial
_{r}\phi\right)  ^{2}+\frac{\lambda_{0}}{2}\phi^{4}\right)  r^{2}dr\nonumber\\
&  =2\pi Q^{2}\int\left(  \left(  r\partial_{r}f-f\right)  ^{2}-\frac{1}%
{2}\alpha_{0}f^{4}\right)  \frac{dr}{r^{2}} \label{61}%
\end{align}
Note that although the contribution of the second term is negative for
negative $\lambda_{0},$ the total energy is positive because the gradient term
dominates. The integral above diverges when $r\rightarrow0$. This divergence
has an ultraviolet origin and it is the same as well known divergence of the
self-energy of classical point-like electric charge. Therefore it can be
removed using standard methods. We will focus instead on the IR contribution
to the energy. Taking into account that at $r\gg R_{c}$,
\begin{equation}
\left\langle \left(  r\partial_{r}f\right)  ^{2}\right\rangle =\left\langle
f^{\prime2}\right\rangle =\alpha_{0}\left\langle f^{4}\right\rangle ,
\label{62}%
\end{equation}
the following expression for the infrared contribution to the total energy is
obtained
\begin{equation}
E\simeq\pi Q^{2}\int_{R_{c}}^{r}\alpha_{0}\left\langle f^{4}\right\rangle
\frac{dr}{r^{2}}\sim O\left(  1\right)  Q^{2}\frac{1}{R_{c}}\left(  \frac
{r}{R_{c}}\right)  ^{1/3}, \label{63}%
\end{equation}
in agreement with the qualitative expectations$.$ The energy of the isolated
charge diverges as $r^{1/3}$ and therefore it cannot exist as a free
\textit{asymptotic state}. This can be interpreted as a hint of confinement of
isolated sources.

One can also build \textquotedblleft\textit{colorless
configuration\textquotedblright} using two opposite charges $Q$ and $-Q$
separated by distance $l$. At distances $\ r\gg l,$ the field $\phi$ decreases
as $r^{-2}$ and equation $\left(  \ref{3}\right)  $ becomes%
\begin{equation}
\frac{1}{r^{2}}\frac{d}{dr}\left(  r^{2}\frac{d\phi}{dr}\right)  =-4\pi
Q\delta\left(  \mathbf{x}\right)  +\lambda_{0}\frac{Q^{3}l^{3}}{r^{6}},
\label{64}%
\end{equation}
Since in this case the anti-screening effect, determined by the last term in
this equation, is completely irrelevant at large distances we conclude that
the total energy of the dipole system is infrared convergent. \ When the
distance between charges exceeds the confinement scale $R_{c}$ the infrared
contribution of the scalar field becomes essential and the total energy is
\begin{equation}
E\sim O\left(  1\right)  R_{c}^{-1}\left(  \frac{l}{R_{c}}\right)  ^{1/3}.
\label{65}%
\end{equation}
Hence, the interaction potential between two charges\ grows as distance in
power one third. This can be interpreted as a confining potential leading to a
natural estimate for the mass scale of the dipole configuration to be of order
$m\sim O\left(  1\right)  R_{c}^{-1}$.

\section{Discussion and Speculations}
We have shown that certain essential properties of the quantum field theory
usually considered as having quantum origin can be revealed already at the
classical level. In particular, the renormalization group structure of the
theory including the phenomenon of dimensional transmutation is already
encoded in the classical equations.

So far we have considered only the self-interacting scalar field with negative
coupling constant and external sources. One can naturally ask up to what
extent the qualitative results obtained in this paper are useful in
application to gauge theories, such as QCD. An encouraging sign, that
classical RG treatment can be generalized for such theories is provided by the
following simple scaling argument. As we have found, the logarithmic effect of
anti-screening comes from \ the term of $\phi^{3}$ in the equation for the
scalar field. In the perturbation theory this term represents the density of
the charge induced by self-interaction and it drops as $r^{-3}$ as distance
$\ r$ grows. In QCD the gauge field equations for gluons contain two kinds of
self-interaction terms which drop in a similar way, namely, $A^{3}$ and
$A\partial A$. Only $A\partial A$ gives the negative contribution to the
$\beta$ function. This term leads to anti-screening effect inducing the
density of the colored charge decaying as $r^{-3}$ similar to the case of
scalar field. Because the structure of the self-interaction terms is different
(in one case it is $\phi^{4}$ and in the other $A^{2}\partial A)$ the
interaction potential between two sources in gauge theories can grow with the
distance not necessarily as $r^{1/3},$ but as $r^{\alpha}$, where
$0<\alpha\leq1.$ So, the linear growth is not excluded. However, the linear
growth, although leading to confined charges does not necessarily imply the
formation of QCD flux tube (see Appendix B).

\bigskip\appendix

\section{On the triviality of $\lambda\phi^{4}$ theory with positive
$\lambda.$}
We can use the exact $\beta$ function equation to check the triviality of
$\lambda_{0}\phi^{4}$ theory in the case of positive $\lambda_{0}$ in four
dimensions (this triviality was rigorously proved in five and higher
dimensions in \cite{frohlich}). In the case of positive $\lambda_{0}$ it is
convenient to change the signs in the definitions of $\alpha_{0}$ and $x$, so
that,%
\begin{equation}
\alpha_{0}\equiv\lambda_{0}Q^{2}>0,\text{ \ \ }x\equiv\ln\left(
r_{0}/r\right)  . \label{67}%
\end{equation}
In this case the $\beta$ function defined in $\left(  \ref{32}\right)  $ is
related to the standard $\beta_{st}$ function used in the literature as
$\beta_{st}=\beta/2.$ With these redefinitions the equation for the exact
$\beta$ function is obtained from $\left(  \ref{49}\right)  $ by substituting
$\beta\rightarrow\beta$ and $\alpha\rightarrow-\alpha:$%
\begin{equation}
\beta=2\alpha^{2}-\frac{1}{2}\left(  \frac{d\beta^{2}}{d\alpha}-\frac
{\beta^{2}}{\alpha}\right)  . \label{68}%
\end{equation}
For $\alpha\ll1$ the perturbative solution of this equation is%
\begin{equation}
\beta=2\alpha^{2}-6\alpha^{3}+48\alpha^{4}-570\alpha^{5}+... \label{69}%
\end{equation}
The Gell-Mann-Low equation to one loop,%
\begin{equation}
\frac{d\alpha\left(  x\right)  }{dx}=2\alpha^{2}\left(  x\right)  , \label{70}%
\end{equation}
gives us%
\begin{equation}
\alpha\left(  r\right)  =\frac{\alpha_{0}}{1-2\alpha_{0}x}=\frac{\lambda
_{0}Q^{2}}{1-2\lambda_{0}Q^{2}\ln\left(  r_{0}/r\right)  }. \label{71}%
\end{equation}
According to this result the coupling constant blows up at the Landau pole%
\begin{equation}
r_{L}=r_{0}e^{-\frac{1}{2\lambda_{0}Q^{2}}}. \label{72}%
\end{equation}
The essence of the proof of $\lambda\phi^{4}$ triviality can be reduced to
showing that this UV pole will survives at nonperturbative level. This is not
so obvious because as we have seen the IR one loop Landau pole (for negative
$\lambda_{0}$) disappears after resummation of the perturbative expansion. To
find out whether the pole survives or not for positive $\lambda,$ let us solve
equation $\left(  \ref{68}\right)  $ in strong coupling regime, $\alpha\gg1.$
We will do it perturbatively in terms of the inverse powers of $\alpha.$
Neglecting the linear $\beta$ term in $\left(  \ref{68}\right)  $ we have%
\begin{equation}
\frac{d\beta^{2}}{d\alpha}-\frac{\beta^{2}}{\alpha}\simeq4\alpha^{2}
\label{73}%
\end{equation}
and the corresponding solution of this equation is%
\begin{equation}
\beta=\sqrt{2}\alpha^{3/2} \label{74}%
\end{equation}
Rewriting $\left(  \ref{68}\right)  $ as%
\begin{equation}
\frac{d\beta^{2}}{d\alpha}-\frac{\beta^{2}}{\alpha}=4\alpha^{2}-2\beta,
\label{75}%
\end{equation}
substituting in the right hand side of this equation the result $\left(
\ref{74}\right)  $ and solving the obtained inhomogeneous linear equation for
$\beta^{2}$ we obtain%
\begin{equation}
\beta=\sqrt{2}\alpha^{3/2}\left(  1-\frac{2\sqrt{2}}{3}\alpha^{-1/2}\right)
^{1/2}=\sqrt{2}\alpha^{3/2}\left(  1-\frac{\sqrt{2}}{3}\frac{1}{\sqrt{\alpha}%
}+O\left(  \left(  \frac{1}{\sqrt{\alpha}}\right)  ^{2}\right)  \right)
\label{76}%
\end{equation}
This procedure can be repeated recursively giving us higher order power
corrections in the expansion in $1/\sqrt{\alpha}\ll1.$ Thus we see that for
very large $\alpha$ the solution $\left(  \ref{74}\right)  $ becomes more and
more accurate and the behavior of $\beta$ functions confirms the expectations
in \cite{LP}$.$ The solution $\left(  \ref{74}\right)  $ exactly matches the
one loop result in $\left(  \ref{69}\right)  $ at $\alpha=1/2.$ Using this to
fix the integration constant in the Gell-Mann-Low equation%
\begin{equation}
\frac{d\alpha\left(  x\right)  }{dx}=\sqrt{2}\alpha^{3/2}, \label{77}%
\end{equation}
we obtain the following non-perturbative result valid for $\alpha\gg1$%
\begin{equation}
\alpha\left(  r\right)  =8\left(  \frac{\alpha_{0}}{1-2\alpha_{0}\left(
x-1\right)  }\right)  ^{2}=8\left(  \frac{\lambda_{0}Q^{2}}{1-2\lambda
_{0}Q^{2}\ln\left(  r_{0}/re\right)  }\right)  ^{2} \label{78}%
\end{equation}
Thus we see that even after complete resummation of the perturbative expansion
the Landau pole survives and its non-perturbative location is at (Fig.\ \ref{fig:Fig3}):%
\begin{figure}[hbt]
\centering
\psfrag{x}{$x=\ln \frac{r_{0}}{r}$}\psfrag{y}{$\alpha(x)$}\psfrag{0.5}[cr][cr]{$\frac{1}{2}$}\psfrag{rl}[tc][cc]{$r_{L}$}\psfrag{rll}[tc][cc]{$\frac{r_{L}}{\mathrm{e}}$}\psfrag{pert}{\begin{tabular}{l}perturbative\\ pole\end{tabular}}\psfrag{nonpert}{\begin{tabular}{l}nonperturbative\\ pole\end{tabular}}
\includegraphics[width=0.8\textwidth]{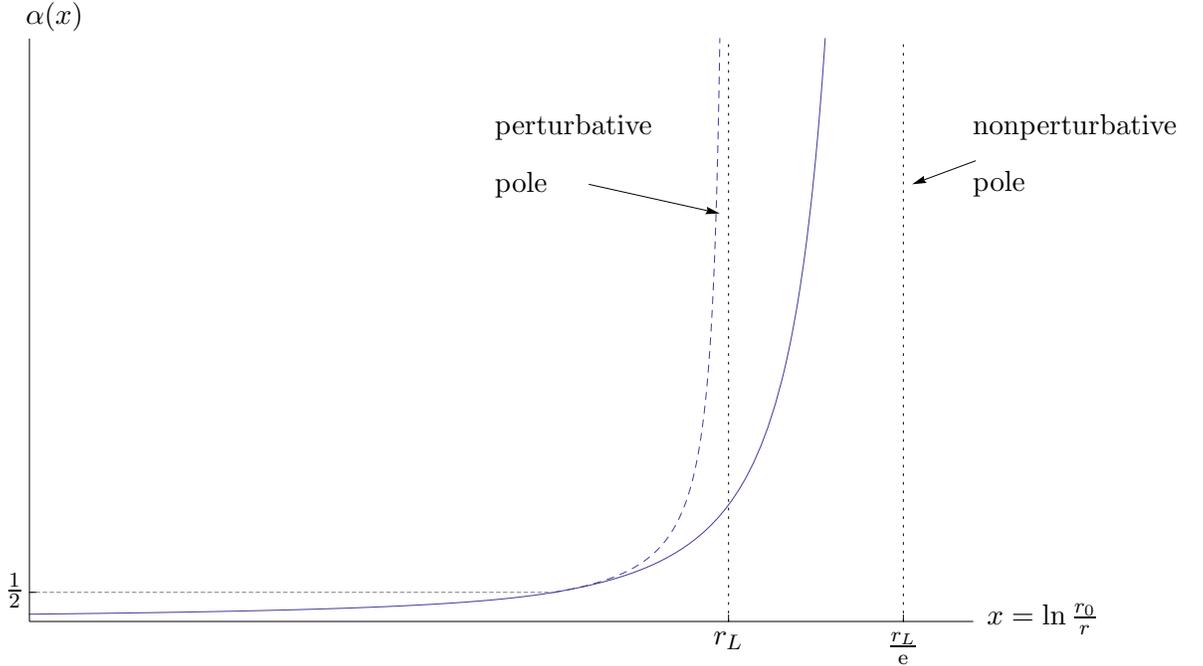}
\caption{Landau pole}
\label{fig:Fig3}
\end{figure}
\begin{equation}
r_{L}=\frac{r_{0}}{e}e^{-\frac{1}{2\lambda_{0}Q^{2}}}. \label{79}%
\end{equation}
This leads to the triviality of $\lambda\phi^{4}$ theory.

\section{Confinement with and without strings$.$}
As we have shown in the paper the potential between the external sources grows
unbounded with the separation as a third power of it. In such a picture,
charge and anti-charge are confined but the confinement in not due to
formation of a flux tube (string-type object) but rather due to formation of a
finite energy dipole. One may think, that this is a peculiarity of not having
a linear growth of the potential. In this appendix we will discuss this issue
and show that even the linear potential does not necessarily imply the
existence of a string.

To demonstrate this we begin with an $SO(3)$ sigma model of an isotriplet
scalar field $\phi^{a}$ $\left(  a=1,2,3\right)  $ with Lagrangian:
\begin{equation}
\mathcal{L}\,=\partial_{\mu}\phi^{a}\partial^{\mu}\phi^{a}\,-\lambda
^{2}\,(\phi^{a}\phi^{a}\,-v^{2})^{2}.\label{lagran}%
\end{equation}
In this case the equation of motion has the static spherically-symmetric
solution (see e.g. \cite{alex})
\begin{equation}
\phi^{a}\,=\,f(r){\frac{x^{a}}{r},}\label{solution}%
\end{equation}
where $x^{a}$ are Cartesian space coordinates and $f(r)$ is the function with
the following asymptotic properties
\begin{equation}
f(0)\,=\,0,~~~f(r)|_{r\gg\,(\lambda v)^{-1}}\,\rightarrow\,v\label{solution1}%
\end{equation}
The size of the region where $f(r)$ is different from $v$ is of order
$r_{c}\,=\,(\lambda v)^{-1}$. The solution above is the 't Hooft - Polyakov
monopole in the limit of zero gauge coupling. Because in this limit the gauge
fields become massless and decouple from $\phi$ this solution is often
referred to as a global magnetic monopole.

\begin{figure}[hbt]
\centering
\psfrag{Q1}[cc][cc]{+Q}\psfrag{Q2}[cc][cc]{-Q}\includegraphics[width=0.6\textwidth]{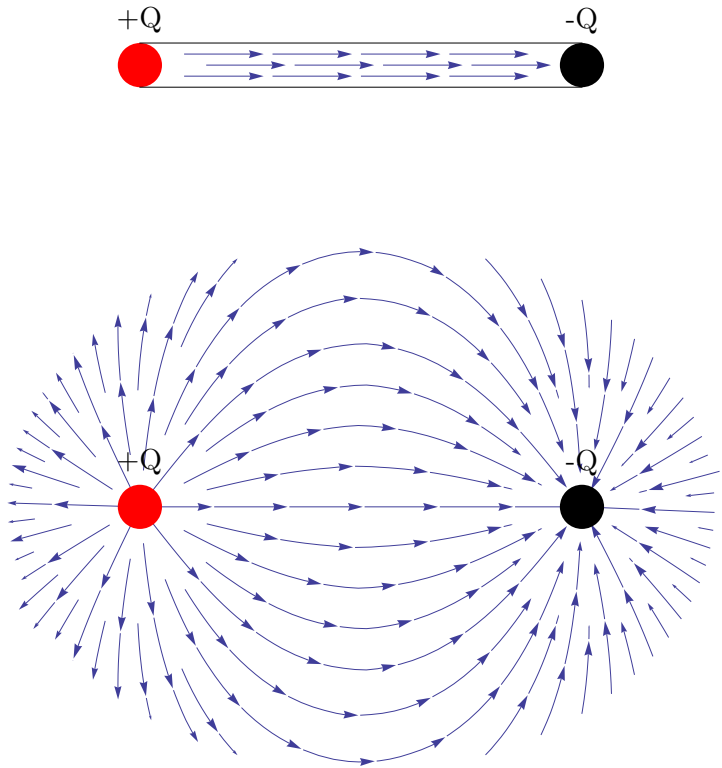}
\caption{String vs.\ dipole}
\label{fig:Fig4}
\end{figure}

The energy of this monopole can be easily estimated by considering separately
the contribution from core $(r<r_{c})$ and from the rest. The core
contribution is of order $v$, and can be neglected compared to the energy in
the region $r>r_{c}$, where $f(r)$ can be set to be equal $v.$ Then the
contribution of the gradients of the angles (Nambu-Goldstone modes) is
divergent and we have to cut-off the integral at some $R\,\gg\,r_{c}$. The
resulting energy of the isolated \textquotedblleft charge\textquotedblright,
\begin{equation}
E_{r\,>\,r_{c}}\,\simeq\,4\pi\int_{r_{c}}^{R}drv^{2}\,\simeq\,4\pi Rv^{2}\,,
\label{energy}%
\end{equation}
is linearly divergent. This implies that the potential between two opposite
charges is linear. In fact, let us consider an anti-monopole placed at
distance $R$ from the monopole. The effect of anti-monopole is to cut the
divergent integral at $r=R$ and the resulting potential is
\begin{equation}
V_{R\,>\,r_{c}}\,\simeq\,R/r_{c}^{2}\,, \label{potmantim}%
\end{equation}
thus confining monopole-antimonopole configuration! This picture is very
different from the QCD flux-tube (string) confinement. To understand this
difference let us confront them considering heavy quark-anti-quark pair placed
at distance $R$ apart. In the absence of light quarks this distance $R$ can be
much larger than the QCD scale, $r_{QCD}\equiv\,\Lambda_{QCD}^{-1}$. In string
picture the force between this pair is mediated by a stretched string
(electric flux tube) of constant tension $\sim\Lambda_{QCD}^{2},$ giving the
potential
\begin{equation}
V_{R\,>\,r_{QCD}}\,\simeq\,R/r_{QCD}^{2}\,, \label{potqantiq}%
\end{equation}
similar to the monopole-anti-monopole potential. However, in the monopole case
the flux is not confined to a string and for monopole-anti-monopole it has a
dipole configuration (see Fig. \ref{fig:Fig4}). 
As a result the \textquotedblleft monopole
color\textquotedblright\ is not bounded within the string of width $r_{c}$ and
can be probed everywhere in the space around the monopole at $r>r_{c}$. On the
other hand, in the case of string the only possibility to probe the color of
the charge is to penetrate within the string of width $r_{QCD}$.

The difference between these two pictures can be stressed even more if we
notice that the theory $\left(  \ref{lagran}\right)  $ allows to be
\textquotedblleft deformed\textquotedblright\ to the theory in which true
strings connecting monopoles appear. The appearance of \textquotedblleft
open-color\textquotedblright\ monopole-antimonopole configuration is due to
$S(3)/U(1)$ topology of the vacuum manifold with nontrivial $\pi_{2}$. To
change this topology we can further deform the vacuum manifold by spontaneous
breaking of the remaining $U(1)$ symmetry. This can be done by introducing
additional scalar field $\chi_{\alpha}$ $\left(  \alpha=1,2\right)  $ in a
doublet representation of the $SO(3)$ group. The Lagrangian then becomes
\begin{equation}
\mathcal{L}\,=\,\partial_{\mu}\phi\partial^{\mu}\phi\,+\partial_{\mu}%
\chi^{\ast}\partial^{\mu}\chi\,-\lambda^{2}\,(\phi^{2}\,-v^{2})^{2}%
-\lambda_{1}^{2}\,(\chi^{\ast}\chi\,-v_{1}^{2})^{2}+\,h\,\chi^{\ast}\phi
\chi\,+h^{\prime\ast}\phi^{2}\chi\,,\label{lagran2}%
\end{equation}
where the contraction of indices is obvious. The parameters ($\lambda
,\lambda_{1},v,v_{1},h,h^{\prime}$) are chosen in such a way that the field
$\chi$ develops an expectation value $v^{\prime}\ll v$. In this limit, core of
the monopole remains nearly unchanged. However at very large distances the
field $\chi$ dramatically changes the monopole field. The presence of the
second field with nonzero expectation value leads to the following
hierarchical symmetry-breaking pattern,
\begin{equation}
SO(3)\rightarrow\,U(1)\,\rightarrow\,1\,,\label{pattern}%
\end{equation}
and the vacuum manifold becomes topologically-trivial. As a result, the static
isolated monopoles can not exist anymore. However, due the hierarchy of
symmetry breaking, $v\gg v^{\prime}$, monopoles do not simply disappear from
the spectrum, but rather get connected by the strings. Either strings or
monopoles are stable in two limits: $v=\infty$ and finite $v^{\prime}$ or $v$
finite and$~v^{\prime}=0,$ respectively. However, when both $v$ and
$v^{\prime}$ are finite, they can only exist as hybrid configuration, namely,
monopoles connected by strings. This picture is more close to the usual string
confinement because here the monopole magnetic flux gets confined into the
string. The thickness of this string is $\sim1/v^{\prime}\,\gg\,r_{c}$, and
its tension is $v^{^{\prime}2}\ln L,$ where $L$ is the string's length. In
other words, the tension of the string is logarithmically divergent. For
example, for the string oriented in $z$ direction the field $\chi$ near the
string but far away from the monopoles, is
\begin{equation}
\chi_{\alpha}\,\simeq\,\delta_{\alpha}^{1}f\left(  \rho\right)  e^{i\theta
}\,,\label{string}%
\end{equation}
where $f(\rho)$ vanishes at $\rho=0$ and approaches constant for
$\rho\,>\,1/v^{\prime}$ in the cylindric coordinates $\rho,\theta$. The energy
of this configuration is logarithmically divergent with the natural cut-off
scale of order string size.

The picture above can be summarized as follows. For $R\,\ll\,1/v^{\prime}$,
the potential between monopoles is linear and field configuration is of dipole
type, but for $R\,\gg\,1/v^{\prime}$ the flux is not spread anymore and
becomes confined by a string. This leads to the modification of the potential
(\ref{potmantim}), which becomes
\begin{equation}
V(R)\sim v{^{\prime2}}R\ln R\,.
\end{equation}
This is not such a dramatic change in the potential, but more important is the
qualitative change of the physical picture, because now the monopoles are
becoming confined by the string. The consideration above illustrates a very
important point, that the confinement can have very distinct physical origin
for the same growing potential between charges.

Finally, we will consider here one more theory, where confining potential is
due to usual electric flux. Let us consider $U(1)$ theory with Lagrangian
\begin{equation}
L\,=\,(F_{\mu\nu}F^{\mu\nu})^{\alpha}\,+\,A_{\mu}j^{\mu}\,.
\end{equation}
Unlike $\left(  \ref{lagran}\right)  ,$ which describes healthy theory, the
legitimacy of this theory as a quantum field theory of $U\left(  1\right)  $
gauge field \ is much less obvious. However, since we are interested only in
geometric properties of the classical electric fluxes, we will use it for our
purposes. In this case the equations of motion are
\begin{equation}
\partial^{\mu}\left(  2\,\alpha(F^{2})^{\alpha-1}F_{\mu\nu}\right)
\,=\,j_{\nu}\,.
\end{equation}
For the static charge $j_{\mu}\,=\,\delta_{\mu}^{0}\delta(r)Q$, which produces
spherically-symmetric electric field,%
\begin{equation}
F_{j0}\,\equiv E_{j}(r)=E(r){\frac{x_{j}}{r},}%
\end{equation}
they become,
\begin{equation}
\partial^{j}\left(  2\,\alpha E(r)^{2\alpha-2}E_{j}(r)\right)  \,=\,\delta
(r)Q.\,
\end{equation}
It immediately follows from here that
\begin{equation}
2\,\alpha E(r)^{2\alpha-1}\,=\,{\frac{Q}{r^{2}}}\,,
\end{equation}
and hence
\begin{equation}
E(r)\,=\,\left(  {\frac{Q}{2\alpha r^{2}}}\right)  ^{{\frac{1}{2\alpha-1}}}\,.
\end{equation}
The energy of an isolated charge smeared over a sphere $r_{0}$ diverges as
\begin{equation}
E_{charge}\,=\,\int_{r_{0}}^{R}r^{2}dr\left(  {\frac{Q}{2\alpha r^{2}}%
}\right)  ^{{\frac{2\alpha}{2\alpha-1}}}\,\sim\,Q^{{\frac{2\alpha}{2\alpha-1}%
}}\,R^{{\frac{2\alpha-3}{2\alpha-1}}}\,,
\end{equation}
for either $\alpha>3/2$ or $\alpha<1/2$ when exponent is positive. Thus, in
both these cases the energy of an isolated charge diverges. However, \ in this
case the finite energy of the charge-anticharge configuration is not for
granted automatically! The situation is much more subtle than in the usual
case because of very strong non-linearity, which make superposition principle
not applicable to $E_{j}$. However, the superposition principle in this case
is valid for $E(r)^{2\alpha-2}E_{j}(r)$. Therefore, the electric field of the
dipole of size $D$ is given by,
\begin{equation}
E(r)_{D}\,\simeq\,\left(  {\frac{QD}{2\alpha r^{3}}}\right)  ^{{\frac
{1}{2\alpha-1}}}\,.
\end{equation}
The corresponding energy is
\begin{equation}
E_{charge}\,\sim\,(DQ)^{{\frac{2\alpha}{2\alpha-1}}}\,R^{{\frac{-3}{2\alpha
-1}}}\,,
\end{equation}
and it is finite only for $\alpha>3/2$.

\bigskip

%\begin{acknowledgement}

\centerline{\bf Acknowledgments}
We are grateful to L. Alvarez-Gaume, C. Bachas, A. Barvinski, M. Henneaux and
I. Sachs for discussions and valuable comments. We would like to thank T.
Hofbaur for the help with preparing figures.

The work of G.D. was supported in part by Humboldt Foundation under Alexander
von Humboldt Professorship, by European Commission under the ERC advanced
grant 226371, by David and Lucile Packard Foundation Fellowship for Science
and Engineering and by the NSF grant PHY-0758032. The work of C.G. was
supported in part by Grants: FPA 2009-07908, CPAN (CSD2007-00042) and HEPHACOS
P-ESP00346. V.M. is supported by TRR 33 \textquotedblleft The Dark
Universe\textquotedblright\ and the Cluster of Excellence EXC 153
\textquotedblleft Origin and Structure of the Universe\textquotedblright.

%\end{acknowledgement}


\begin{thebibliography}{9}                                                                                                %

\bibitem {Gross}Gross, D., Wilczek, F. Ultraviolet behavior of non-Abelian
gauge theories. \textit{Phys. Rev. Lett.}, \textbf{30} (1973), 1343; Politzer,
H. Reliable perturbative results for strong interactions? \textit{Phys. Rev.
Lett.}, \textbf{30} (1973), 1346.

\bibitem {simanzek}Symanzik, K. A field theory with computable large momenta
behavior. \textit{Lett. al Nuovo Cimento}, \textbf{6} (1973), 77; Parisi, G.
Deep inelastic scattering in a field theory with computable large-momenta
behavior. \textit{Lett. al Nuovo Cimento}, \textbf{7} (1973), 84.

\bibitem {wilson}Wilson, K. RG and strong interactions. \textit{Physical
Review}, \textbf{D 3} (1971), 1300; Wilson, K., Kogut, J. The renormalization
group and the $\varepsilon$-expansion. \textit{Phys. Rep}. 12 (1974), 76; 't
Hooft, G. \textit{Nucl. Phys}. \textbf{B61} (1973) 455 ; The renormalization
group and Quantum Field Theory, Doorworth, 1988.

\bibitem {GML}Gell-Mann, M., Low, F. E. Quantum Electrodynamics at Small
Distances. \textit{Physical Review}, \textbf{95} (5) (1954), 1300.

\bibitem {landau}Landau, L. D., Abrikosov, A. A., and Khalatnikov, I. M.
\textit{Dokl. Akad. Nauk SSSR,} \textbf{95}, 497, 773, 1177 (1954).

\bibitem {lipatov}Lipatov, L. N. \textit{Zh.Eksp.Teor.Fiz.} \textbf{72} (1977)
411, [\textit{Sov.Phys. JETP} \textbf{45} (1977) 216]; Brezin, E., Le Guillou,
J.-C., Zinn-Justin, J. Perturbation theory at large order \textit{Phys. Rev.
}\textbf{D 15} (1977), 1544, 1558; Parisi, G. \textit{Phys.Lett.} \textbf{66B}
(1977) 382.

\bibitem {LP}Landau, L. D., Pomeranchuk, I. Ya., \textit{Dokl. Akad. Nauk
SSSR,} \textbf{102} (1955) 489; Pomeranchuk, I. Ya. \textit{Dokl. Akad. Nauk
SSSR,} \textbf{103} (1955) 1005.

\bibitem {frohlich}Aizenman, M. Geometric analysis of $\phi^{4}$ fields and
Ising models, I, II. \textit{Comm. Math. Phys.} 86 (1982) 1; Froehlich, J. On
the triviality of $\lambda\phi^{4}$ theories and the approach to the critical
point in d(-)
%TCIMACRO{\TEXTsymbol{>} }%
%BeginExpansion
$>$
%EndExpansion
4 dimensions, \textit{Nuclear Phys.} \textbf{B 200 }(1982) 281.

\bibitem {alex}Vilenkin A., Shellard E.P.S. Cosmic strings and other
topological defects. Cambridge University Press, 1994
\end{thebibliography}
\end{document}